\documentclass[aps,10pt,pra,twocolumn,showpacs,showkeys,groupedaddress]{revtex4-1}
\usepackage[colorlinks,linkcolor=red,citecolor=blue,urlcolor=blue]{hyperref}
\usepackage{amsmath}
\usepackage{graphicx}
\usepackage{bm}
\usepackage{xcolor}
\usepackage{comment}
\usepackage[normalem]{ulem}
\usepackage{enumitem}
\usepackage{natbib} 
\usepackage{microtype}

\DeclareBoldMathCommand{\bfmu}{\mu}
\newcommand{\be}{\begin{equation}}
\newcommand{\ee}{\end{equation}}
\newcommand{\bea}{\begin{eqnarray}}
\newcommand{\eea}{\end{eqnarray}}
\def\mum{\rm\mu m}
\def\mus{\rm\mu s}

\def\muK{\rm\mu K}

\def\ie{{\it i.e.,\/}}
\def\eg{{\it e.g.,\/}}
\def\vs{{\it vs.\/}}

\def\via{{\it via\/}}

\def\insitu{{\it in~situ\/}}

\def\GP{Gross-Pitaevskii}

\def\negskip{\vskip-\baselineskip}
\begin{document}

\title{Robust spatial coherence $\mathbf{5\,\mu m}$ from a room-temperature atom chip}

\author{Shuyu Zhou}	
	\thanks{\footnotesize Present address: Key Laboratory for Quantum Optics, Shanghai Institute of Optics and Fine Mechanics, The Chinese Academy of Sciences, Shanghai 201800, China.}
	\affiliation{Department of Physics, Ben-Gurion University of the Negev, Be'er Sheva 84105, Israel}
\author{David Groswasser}
	\affiliation{Department of Physics, Ben-Gurion University of the Negev, Be'er Sheva 84105, Israel}	
\author{Mark Keil}
	\thanks{\footnotesize Corresponding authors:
	\href{mailto:mhkeil@gmail.com}{\tt mhkeil@gmail.com}, \href{mailto:folman@bgu.ac.il}{\tt folman@bgu.ac.il}
	\\}
	\affiliation{Department of Physics, Ben-Gurion University of the Negev, Be'er Sheva 84105, Israel}	
\author{Yonathan Japha}
	\affiliation{Department of Physics, Ben-Gurion University of the Negev, Be'er Sheva 84105, Israel}	
\author{Ron Folman}
	\thanks{\footnotesize Corresponding authors:
	\href{mailto:mhkeil@gmail.com}{\tt mhkeil@gmail.com}, \href{mailto:folman@bgu.ac.il}{\tt folman@bgu.ac.il}
	\\}
	\affiliation{Department of Physics, Ben-Gurion University of the Negev, Be'er Sheva 84105, Israel}

\begin{abstract}
	
We study spatial coherence near a classical environment by loading a Bose-Einstein condensate into a magnetic lattice potential and observing diffraction. Even very close to a surface ($\rm5\,\mum$), and even when the surface is at room temperature, spatial coherence persists for a relatively long time ($\rm\ge500\,ms$). In addition, the observed spatial coherence extends over several lattice sites, a significantly greater distance than the atom-surface separation. This opens the door for atomic circuits, and may help elucidate the interplay between spatial dephasing, inter-atomic interactions, and external noise.

\end{abstract}

\date{\today}

\maketitle

\section{Introduction\label{sec:introduction}}

Recent developments in ultracold atomic physics include realistic prospects for atomic circuits~\cite{Pepino2009,Charron2006,Japha2007,Ryu2013,Eckel2014,NJP_SpecialIssue,Caliga2016}. Such an analogue of electronics, which has been coined ``atomtronics'', offers significant opportunities in both fundamental and applied physics, due to the rich atomic degrees of freedom. To fully realize such a device, at least three milestones~--~adapted from electronic devices~--~need to be met: the ability to design and realize arbitrary potentials for guides and traps; single-site addressability; and controllable interactions and transport \via\ tunneling barriers. The latter requirement demands that potentials be sculptured with a resolution on the order of the de-Broglie wavelength of the atoms or less ($<\rm1\,\mum$). To achieve these milestones in a scalable, ``solid state''-type of device, one must be able to trap the atoms and manipulate their external degrees of freedom coherently not more than a few micrometers away from the surface used to generate the potential fields~\cite{Salem2010}. 

Atom chips~\cite{Folman2002,Reichel2002,Fortagh2007} are promising candidates as a platform for atomic circuits. Alongside fundamental experiments~\cite{Obrecht2007a,Maussang2010,Riedel2010,vanZoest2010,Langen2013,Margalit2015}, impressive progress towards clocks, acceleration sensors and quantum information processing is being made on these chips.

Interference or diffraction patterns, hallmarks of spatial coherence, have so far been observed for trapped atoms only when held~$\sim\rm50\,\mum$ or further from atom chip surfaces~\cite{Wang2005,Schumm2005a,Jo2007,Baumgartner2010,Berrada2013} (in contrast, internal-state coherence has been realized much closer to the surface~\cite{Treutlein2004}). These experiments were not intended to measure (or avoid) disruptive effects close to surfaces, and their observed loss of coherence is at least partially attributable to atom-atom interactions. Additional disruptive influences that may have affected previous attempts to measure spatial coherence below~$\sim\rm10\,\mum$ from the surface include potential corrugations, Johnson noise, technical current fluctuations, quasi-condensate phase fluctuations, and heating effects~(\cite{Keil2016x} and references therein). We note that spatial coherence has also been demonstrated~$\rm44\,\mum$ from an atom chip using trapped-BEC interferometry with internal-state labeling~\cite{Bohi2009}.

Permanent-magnet atom chips have been used successfully to trap atoms~6-$\rm8\,\mum$ from the surface but interference between adjacent sites was not observed~\cite{Leung2014,Jose2014,Surendran2015}, while diffraction has been observed for atoms dynamically reflected from surfaces, but without trapping~\cite{Bender2014}. The present work combines long-lived magnetic lattice traps with measurements of diffraction when the atoms are released, thus demonstrating persistent spatial coherence very close~($\rm5\,\mum$) to the atom chip surface.

Even when the surface is at room temperature (thus creating an extreme temperature differential from the ultracold atoms), we find that spatial coherence over a length of at least~15$\rm\,\mum$ can be maintained for a relatively long time,~$\tau_{\rm coh}\ge\rm500\,ms$. Specifically, we enter the regime in which the distance of the atoms from the classical environment of an atom chip is smaller than the observed coherence length. In this regime, in which the correlation length of the Johnson noise is smaller than the probed coherence length, dephasing should be strongest. Furthermore, it is apparent that potential corrugations due to material impurities and fabrication imperfections that become evident close to the surface~\cite{Wang2004,Japha2008,Aigner2008} do not destroy the long-lived spatial coherence achieved in this study. The experiment described here thus constitutes a major step towards the realization of atomtronics with atom chips.

Section~\ref{sec:experimental} of this paper outlines the loading of the~BEC into the magnetic lattice of our atom chip, as well as our data analysis procedures. In Sec.~\ref{sec:results} we show that the fringe patterns observed upon release from the magnetic lattice are indeed deterministic, exhibiting repeatable fringe positions and spacings indicative of phase preservation. Our analysis of these fringes is conducted in Sec.~\ref{sec:analysis} in terms of the spatial coherence length. The evolution of the atomic cloud while trapped in the magnetic lattice, and its subsequent expansion after release, are calculated in the \GP\ approximation as described in Sec.~\ref{sec:theory}. These calculations enable a quantitative understanding of factors affecting the experimental results and provide a basis for further discussion in Sec.~\ref{sec:discussion}. Finally, we summarize our results and conclude in~Sec.~\ref{sec:summary}.

\section{Experimental procedures\label{sec:experimental}}

\subsection{Magnetic lattice loading and release\label{subsec:setup}}

\begin{figure}[t!]
      \centering
      \includegraphics[width=0.5\textwidth]{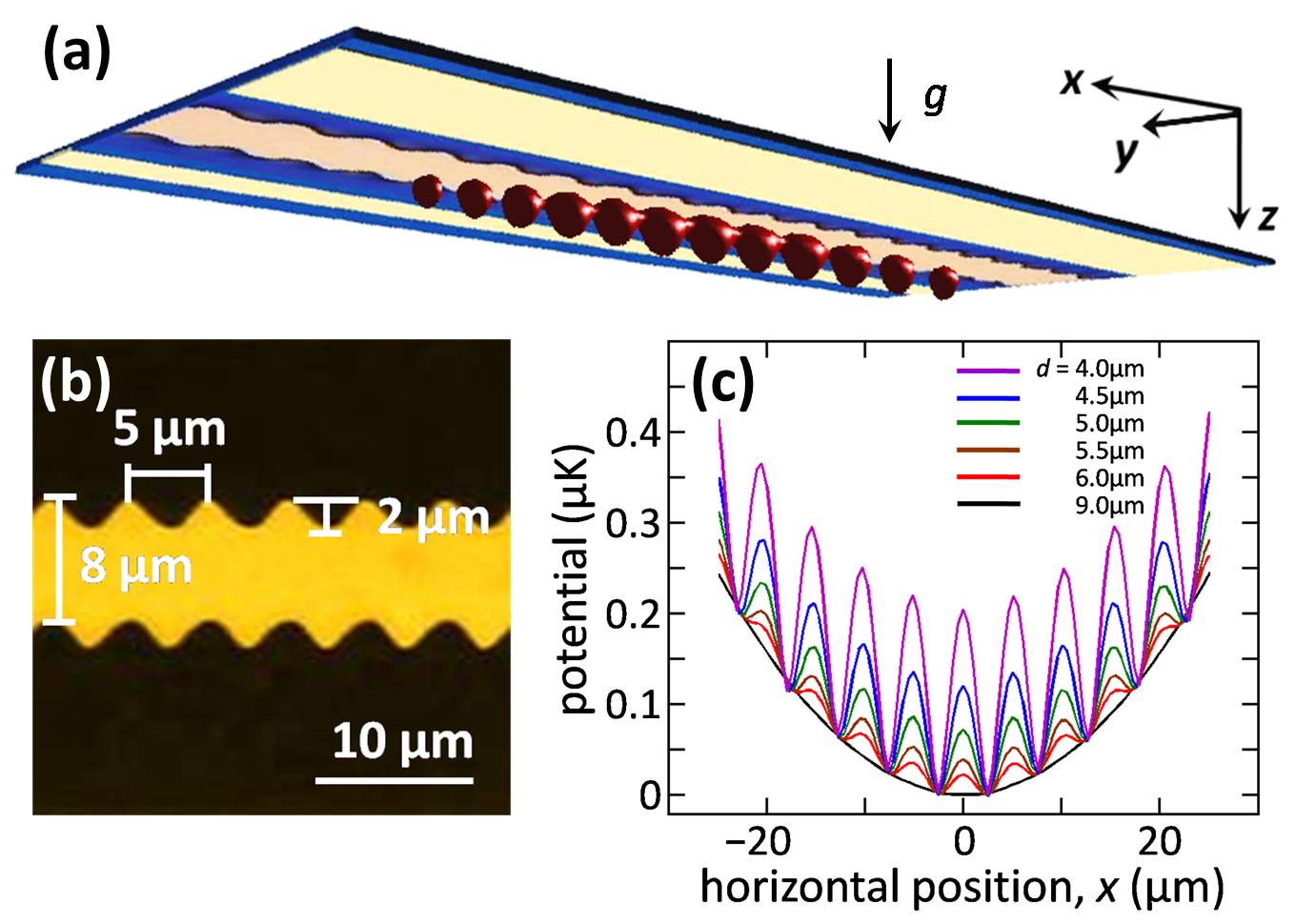}	
   \caption{(Color online) Experimental configuration. (a)~An artist's view of the trapped cloud of atoms a few~$\mum$ from the surface of an atom chip. The atoms are trapped below the surface to allow their spatial distribution to evolve after release from the trap without being adsorbed onto the chip. (b)~An optical microscope image of the current-carrying ``snake wire'' ($\rm500\,nm$-thick Au) that creates a magnetic lattice potential along the~$x$ axis. The main ``trapping wire'' is not shown. (c)~The modulated potential along the~$x$ axis is depicted together with the harmonic potential produced by the trapping wire, giving rise to the atom density profile sketched in~(a). The trapping wire is used to adjust the distance~$d$ of the trap from the snake wire, with progressively stronger potential modulation controlled by reducing~$d$ from~$\rm9.0\,\mum$ (no modulation) to~$\rm4.0\,\mum$ (greatest modulation shown). The modulation amplitude may also be fine-tuned by adjusting the snake-wire current. Not shown is the radial confinement potential that prevents the atoms from hitting the surface. The imaging beam propagates along the~$y$ axis but we do not provide \insitu\ images of the modulated atom density because of insufficient imaging resolution ($\sim\rm7\,\mum$).}
   \label{fig:schematic}
\end{figure}

The experiment is conducted using an atom chip setup~\cite{Folman2002,Reichel2002,Fortagh2007}, as shown schematically in Fig.~\ref{fig:schematic}(a). The~BEC is generated by collecting~$\rm^{87}Rb$ atoms in a magneto-optical trap and transferring them, in the~$|F=2,m_F=2\rangle$ hyperfine state, into an elongated magnetic trap created by the current in a large copper Z-shaped wire (the ``trapping wire'') and bias fields in the~$x$ (longitudinal), $y$ (imaging) and~$z$ (vertical) directions. After~RF evaporative cooling, the~BEC contains about~$10^4$ atoms at~$z\approx\rm340\,\mum$ from the atom chip surface.

The sinusoidal shape of the atom chip ``snake wire'' shown in Fig.~\ref{fig:schematic}(b) causes its current to periodically change direction, thereby modulating the magnetic potential with the same~$\rm5\,\mum$-periodicity of the wire. The modulated potential becomes effective for distances closer than~${\approx\rm9\,\mum}$ from the chip surface [Fig.~\ref{fig:schematic}(c)]. 

Loading the magnetic lattice is realized in two steps. The first step brings the condensate to~$z\approx\rm25\,\mum$ by reducing the trapping-wire current. Simultaneously, we reduce the bias field in the~$x$ direction, thereby increasing the magnetic field at the trap minimum from~$\rm0.2\,G$ to~$\rm18.3\,G$. We also turn on the snake wire with a current of~$\rm30\,mA$. At the end of the first step, performed in~$\rm200\,ms$, the~BEC is located in a smooth harmonic trap produced by currents in both the trapping wire and the snake wire, as in previous work~\cite{Zhou2014}.

In the second step, we reduce the current in the snake wire to bring the trap location down to~$z=\rm5.0\pm0.5\,\mum$ (see Appendix~\ref{sec:appendix_calib}). This turns on the modulation [Fig.~\ref{fig:schematic}(c)]. The second step is completed in~$\rm8\,ms$, a much shorter time than required for perfectly adiabatic loading; we carefully optimize a non-linear current ramp in order to avoid oscillation and excitation~\cite{Schaff2001}. Fine adjustments of the trap position in the~$x$, $y$, and~$z$ directions are pre-determined by currents in a pair of~U-shaped atom chip wires~\cite{Zhou2014}, the $z$-axis bias coils, and very slight changes in the trapping-wire current, respectively. 

As shown in Fig.~\ref{fig:schematic}(c), the modulation of the potential may be controlled by adjusting the distance of the trap from the surface. Indeed, experiments conducted with an atomic cloud located a few~$\rm\mum$ further from the surface show no effects due to the potential modulation (Sec.~\ref{subsec:focusing}), while enhanced effects are seen when the cloud is moved closer. 

The external bias field and currents used for the final trap are~${\bf B}_{\rm ext}=\rm(0.0,39.3,0.2)\,G$, $I_{\rm Z}=\rm32.1\,A$, and~$I_{\rm snake}=\rm5.5\,mA$. Isopotentials for this trap show that its depth is~$\rm\approx2\,\muK$, with negligible perturbations from the Casimir-Polder force for~$z\gtrsim\rm2\,\mum$~\cite{Salem2010}.

After a brief holding time in the trap and subsequent release, we measure up to~4000 atoms, indicating some loss upon loading into the trap. We do not measure the number of atoms in the magnetic lattice trap \insitu, since such measurements would only be qualitative due to high magnetic fields and optical densities. 

The field at the trap minimum is kept at a relatively high value of~$B_0=\rm18.3\,G$ in order to avoid generating a trap with an excessively high aspect ratio. This reduces random phase fluctuations that are characteristic of the~1D BEC regime~\cite{Dettmer2001} and also facilitates optimization of the ``launching'' stage implemented just before release (see below). Smaller atom-surface distances could be achieved in future experiments by reducing~$B_0$, which would increase the potential barrier to the surface, while maintaining the aspect ratio through more control over the longitudinal frequency~$\omega_x$.

The~BEC is held for times~$t=\rm30-500\,ms$ in the magnetic lattice at~$z=\rm5.0\pm0.5\,\mum$. The harmonic trap frequencies in the longitudinal and transverse directions are~$\omega_x=\rm2\pi\times45\,Hz$ and~$\omega_y\approx\omega_z\approx\rm2\pi\times950\,Hz$, respectively. The longitudinal frequency is measured for an atom-surface distance~$>\rm9\,\mum$ where there is no potential modulation [Fig.~\ref{fig:schematic}(c)]; this frequency does not depend strongly on the distance from the surface. The transverse harmonic frequencies are calculated since they do depend strongly on the distance. Finally, we estimate that the peak-to-valley potential modulation is~$\approx\rm80\,nK$ at~$z=\rm5.0\,\mum$, with a longitudinal frequency within each magnetic lattice site of about~$\omega_{\rm site}=\rm2\pi\times\rm500\,Hz$. {\it In~situ} absorption images show that the~BEC is up to~$\rm30\,\mum$ long, thereby covering~6 sites of the magnetic lattice.

Releasing the condensate after the holding time~$t$ is conducted in two steps. We first ``launch'' the condensate by suddenly increasing the current in the snake wire from~$\rm5.5\,mA$ to~$\rm18\,mA$ in~$\rm100\,\mus$. The launching step has two functions. First, it forces the atoms away from the chip so that they avoid crashing into the surface, which would otherwise occur as the atomic cloud expands. Second, and more important, the cloud still experiences the harmonic longitudinal confinement potential (the trapping wire current is still turned on), creating a focusing effect for the~BEC~\cite{Castin1996,Shvarchuck2002,vanAmerongen2008}. The trap is fully released~$\rm2.3\,ms$ later by turning off all currents and fields, finally allowing free-fall under gravity. We use resonant absorption imaging to measure the atomic density distribution after~$\rm12\,ms$ of time-of-flight~(TOF). The focusing effect engineered by our two-step release sequence plays a crucial role in the experiment: conducting the complete trap release in a single step would require an impractically long time-of-flight to develop the far-field diffraction pattern that we wish to study~(Sec.~\ref{subsec:focusing}).

\subsection{Data acquisition and averaging\label{subsec:data}} 

\begin{figure}[t!]
   \centering
   \includegraphics[width=0.45\textwidth]{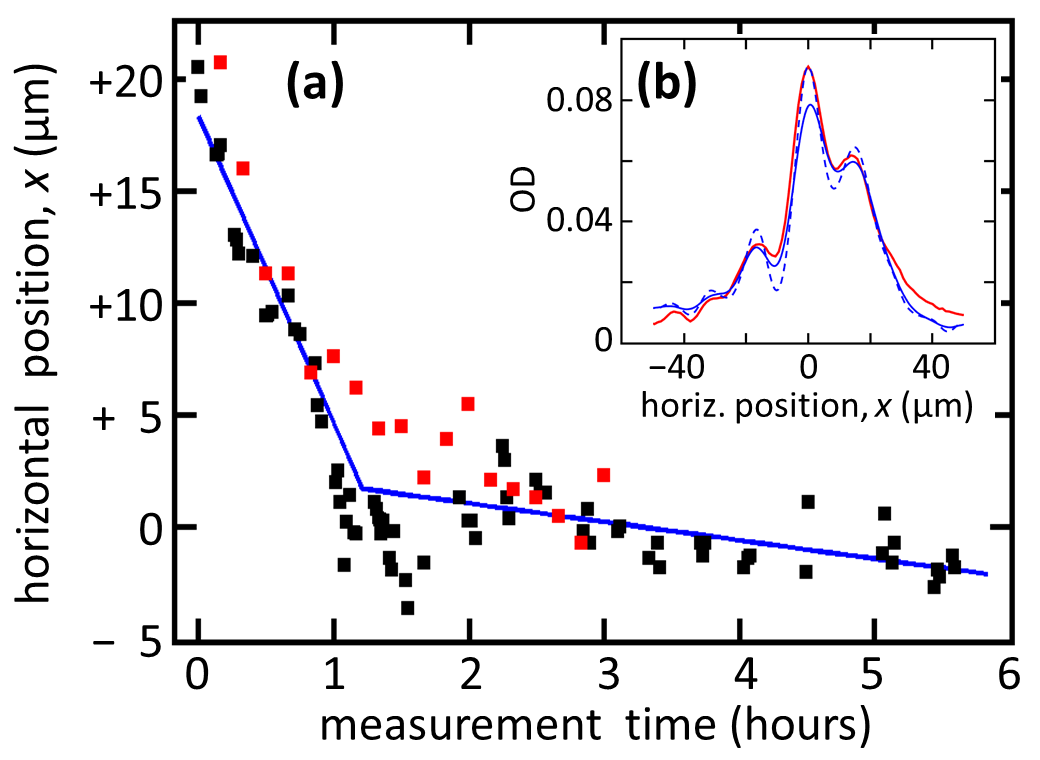}
   \caption{(Color online) Experimental drifts and the effect of jitter. (a)~Horizontal~($x$) position of the center of the atomic cloud, measured~\insitu\ during the first~3-$\rm5\,hr$ of two experimental runs performed several weeks apart (black and red squares). The blue lines depict linear fits to a fast movement during an initial~$\approx1$-hr warm-up period, followed by a slower drift for the remainder of these experimental runs. (b)~Loss of contrast that can be caused by residual ``jitter'' (\ie\ shot-to-shot variability in the \insitu\ horizontal position) and inaccuracies in our post-correction procedures. The blue curves are calculated by applying random displacements along~$x$ to a short-term average of~30 consecutive experimental cycles, with standard deviations of~$\rm2.0\,\mum$ (dashed) and~$\rm3.6\,\mum$ (solid). The extent of smearing due to assumed jitter is consistent with data obtained by averaging over~200 experimental cycles acquired over several days (red curve).} 
   \label{fig:drift}
\end{figure}

Spatial coherence may be demonstrated by repeated realizations of the experiment showing the same location of the spatial interference fringes. Such a demonstration requires a high level of stability with respect to various sources of experimental drift and noise. During long experimental runs that span many days, our atom chip mount is observed to drift slowly along the horizontal~($x$) direction, apparent as motion of the \insitu\ images shown in Fig.~\ref{fig:drift}. Here we describe our post-selection and post-correction procedures, implemented to avoid excessive smearing of the observed fringes that can be caused by such drifts when averaging many experimental realizations. These procedures are not required for short-term data samples however, for which we are able to observe stable fringes without applying any such post-selection or post-correction as shown in Sec.~\ref{sec:results}.

The initial warm-up period of the experiment produces a drift of about~$\rm20\,\mum$ and appears to be caused by thermal stresses in the atom chip mount upon ohmic heating of the copper wires and their leads. The resulting translation of the entire atom chip mount equilibrates after about~1-1$\rm\frac{1}{2}\,hr$ of operation. We do not use data taken during this warm-up period. A much slower drift continues even after the warm-up period and appears to be caused by slight movements in the position of the trapping wire relative to the atom chip itself. As small as these slow-drift movements are however, they are a significant fraction of the magnetic lattice spacing and could cause significant smearing of the observed diffraction pattern.

The~\insitu~$x$-axis position of the condensate depends on the harmonic potential created by the trapping wire. This provides a simple method to correct for these small slow-drift movements after the initial warm-up period.

For a given holding time, we acquire a series of~5-10~TOF images, after which we re-measure the \insitu\ position in an additional experimental cycle. Our post-selection procedure is then straightforward: we reject the entire series of~5-10 images if the \insitu\ horizontal position has wandered outside a range of~$\pm7\,\mum$ or if the distance from the atom chip is outside the range~$z=\rm5.0\pm0.5\,\mum$. These selection criteria generally result in rejecting about~60-70\% of the experimental cycles. We then apply a post-correction procedure to account for the slow drift by shifting each measurement in the series by an average of the shifts observed in the \insitu\ images taken at the beginning and end of the series. These corrections are~$\lesssim\rm3\,\mum$ since this is the maximum difference between successive \insitu\ measurements, evident as the shot-to-shot variability during the slow-drift period in Fig.~\ref{fig:drift}(a). Finally, we average all the post-selected and post-corrected images for a given holding time. Despite various instabilities [\eg\ as depicted in Fig.~\ref{fig:drift}(b)], these averages show stable fringes, even for experiments conducted over several weeks that accumulated over~1000 images (Sec.~\ref{sec:results}).

We find an additional slow drift in the \insitu\ distance of the cloud from the atom chip surface. This is caused by vertical~($z$-axis) motion of the trapping wire due to ohmic heating of its copper leads, and results in a monotonically increasing atom-surface distance. We periodically adjust the trapping-wire current (by~$<2\%$) to maintain this distance within the range noted above and of course, we repeat the \insitu\ calibration measurements before proceeding with further~TOF acquisition.

The shot-to-shot positional variability (``jitter'') evident in Fig.~\ref{fig:drift}(a) cannot be easily compensated, and it inevitably contributes to smearing the fringes that we observe after long-term data acquisition and averaging. Figure~\ref{fig:drift}(b) shows quantitatively how a short-term data sample, convoluted with assumed amounts of random jitter, produces the less well-resolved fringes characteristic of our longer-term averages. The amounts of added jitter assumed in generating these comparisons are consistent with the jitter seen after the initial warm-up period of Fig.~\ref{fig:drift}(a).

\section{Results\label{sec:results}}

\begin{figure}[t!]
   \centering
   \includegraphics[width=0.45\textwidth]{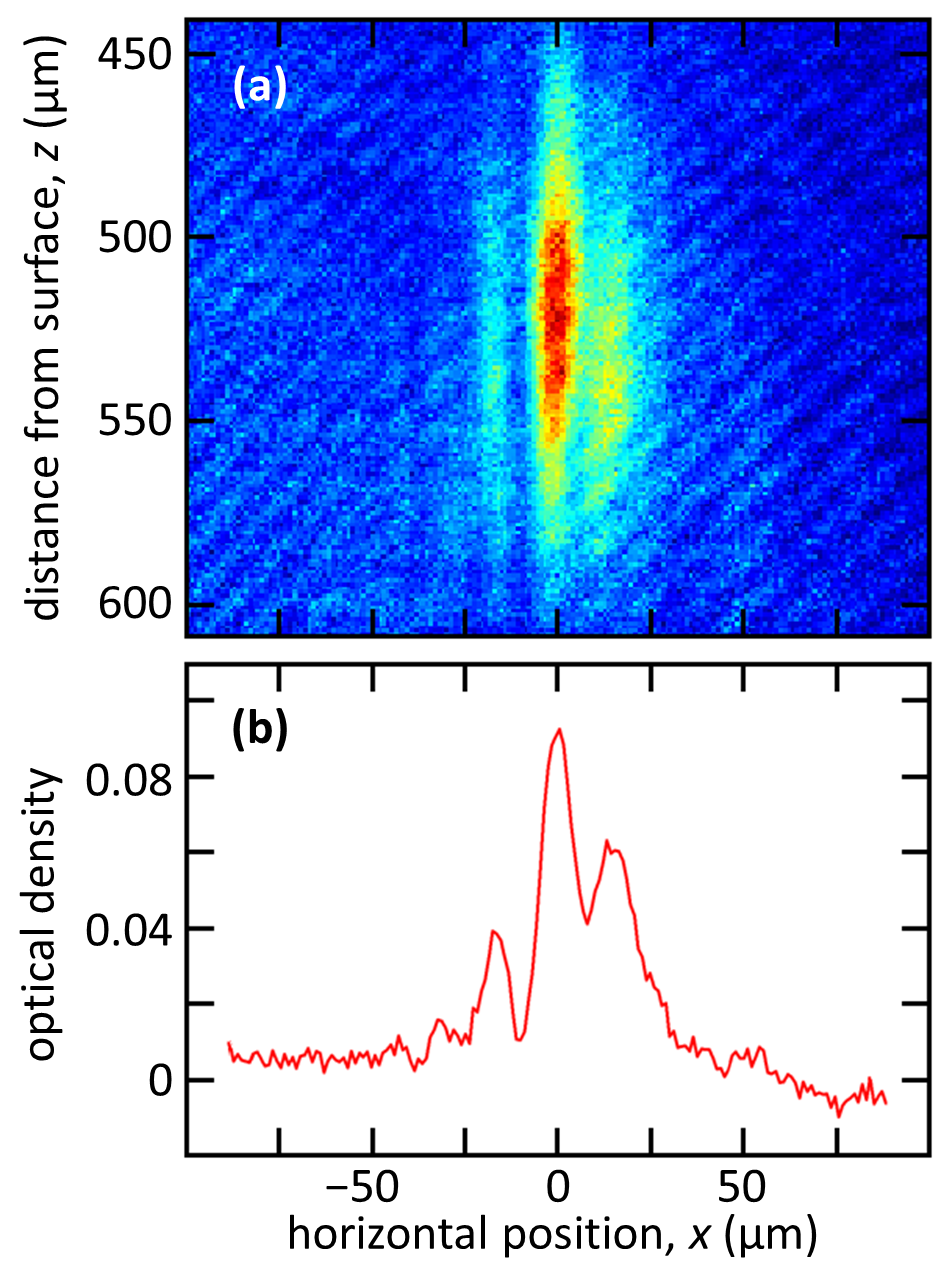}
   \caption{(Color online)
   Experimental signal. Diffraction pattern observed after a trap holding time of~$t=\rm100\,ms$ for~30 consecutive experimental cycles (no post-selection or post-correction is used). (a)~Average of images acquired by absorption imaging. (b)~A cut through the center of~(a). The pronounced contrast of~$\approx0.6$ is highly unlikely to result from any random processes, thus demonstrating spatial coherence. The origin of the observed asymmetry is discussed in Sec.~\ref{subsec:asymmetry}. Only a narrow vertical integration band is required for obtaining a relatively high signal-to-noise ratio, but a wider band would not significantly reduce the observed contrast since the fringes are straight and parallel to the vertical axis~$z$.} 
   \label{fig:30shots}
\end{figure}

An average of~30 consecutive images is presented in Fig.~\ref{fig:30shots}(a-b) for a trap holding time of~$\rm100\,ms$. The zero- and first-order diffraction peaks are clearly visible. For this short-term sample, no post-selection or post-correction has been used. We have confirmed experimentally that the observed diffraction pattern periodicity of about~$\rm15\,\mum$ is independent of the trap-to-surface distance, the purity of the~BEC, the position of the trapped cloud along the lattice, and the amplitude of the diffraction orders. The expected fringe pattern periodicity for a time-of-flight of~$t_{\rm TOF}=\rm12\,ms$ is~$ht_{\rm TOF}/m\lambda=\rm11\,\mum$, where~$\lambda=\rm5\,\mum$ is the lattice period. The launch time of~$\rm2.3\,ms$ also contributes to the fringe separation, so the measurement is only slightly larger than this simple estimate. Our imaging may also contribute to this slight difference (Appendix~\ref{sec:appendix_calib}).

\begin{figure}[t!]
   \includegraphics[width=0.50\textwidth]{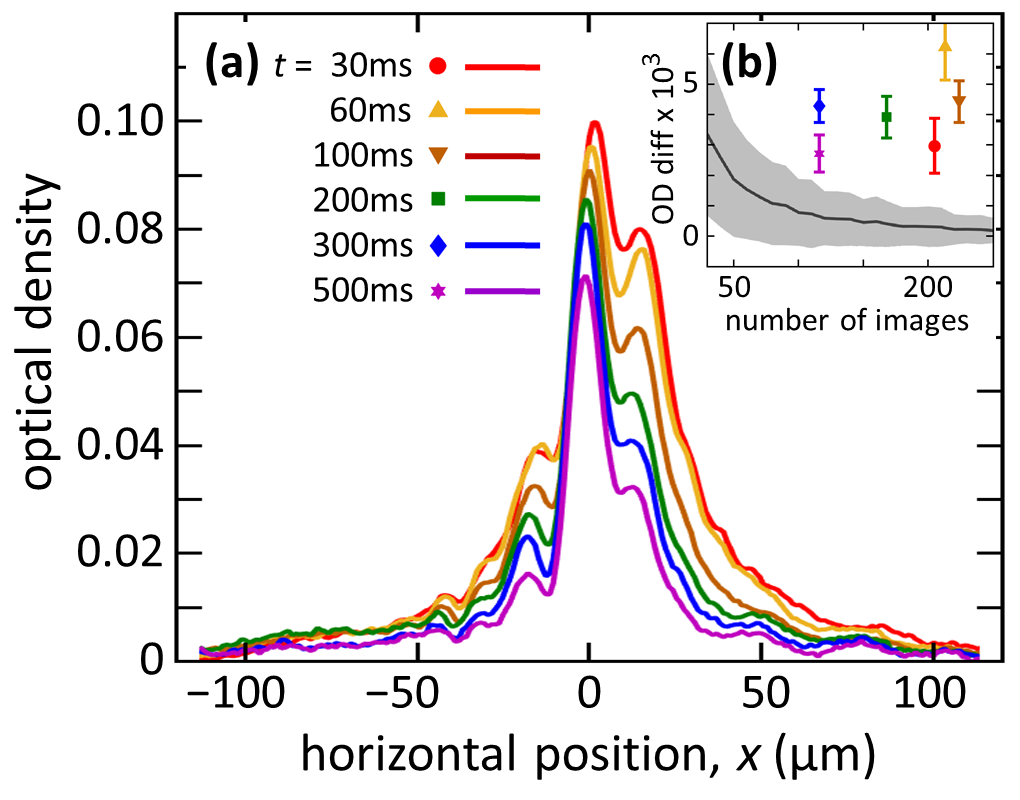}	
   \caption{(Color online)
   Robust spatial coherence. (a)~Repeating the cut of Fig.~\ref{fig:30shots}(b) for trap holding times of \hbox{$t=\rm30-500\,ms$,} averaged over all post-selected experimental cycles for each holding time. Here we include a~$\rm100\,\mum$-wide wide band along the vertical~($z$) axis in order to improve the signal-to-noise ratio. We find that the first-order peaks are ``locked'' at about~$\rm\pm15\,\mum$, independent of~$t$, for the~$>1000$ images collected in this figure. The progressively declining~OD for increasing~$t$ indicates an atom lifetime of~$\rm500\pm50\,ms$, consistent with measured spin-flip rates for this experiment, which are mostly due to technical noise~\cite{Japha2016}. (b)~Data points [same color code as in~(a)] show the optical density difference~(``OD~diff'') between the diffraction side-peak maxima and minima, averaged over all the experimental images obtained for each holding time. Error bars are extracted from a bootstrapping procedure~\cite{Efron1986}. The black curve simulates how the same~OD difference, after averaging a given number of images, would drop towards zero for a random distribution of phases amongst the potential wells. The shaded band around this curve shows a~$1\sigma$ standard deviation for the average that would be caused by such random phases. The data lie~$\rm2-5\,\sigma$ outside this band.}
   \label{fig:1000shots}
\end{figure}
   
The observed contrast is much higher than would be expected from an average of~30 images with random fringes, thus demonstrating coherence~\cite{Hadzibabic2004}, and is typical of the superfluid phase of a~BEC~\cite{Greiner2002}. These results show that the~BEC's most fragile feature, its spatial coherence, can be maintained for at least as long as, for example, spin-flip lifetimes for neutral atoms, even so close to a room-temperature surface.

A much larger sample of images~--~more than~1000~--~is presented in Fig.~\ref{fig:1000shots}(a). This sample includes holding times~$t$ up to~$\rm500\,ms$, much longer than in Fig.~\ref{fig:30shots}. We find that the first-order diffraction peaks are ``locked'' at about~$\pm15\,\mum$, independent of~$t$, clearly demonstrating the robustness of the observed fringes. Nevertheless, the fringe contrast is reduced compared to Fig.~\ref{fig:30shots}(b). This loss of contrast may arise in several ways. Experimental factors include detection noise due to low optical density, thermal background, imperfect loading of the magnetic lattice that may populate excited states, and shot-to-shot positional jitter [Fig.~\ref{fig:drift}(b)]. We do not compensate for these factors in the data analysis. We do, however, compensate for long-term drifts: due to the large number of experimental cycles used here (spanning several days for each holding time), slight drifts of the lattice and experimental conditions become important. The data in Fig.~\ref{fig:1000shots} are therefore post-selected and post-corrected for a range of atom-surface distances and horizontal placements of the trapped cloud, as discussed in Sec.~\ref{subsec:data}. 

Factors that may cause intrinsic loss of contrast include~1D quasi-condensate phase fluctuations that would produce far-field peaks with random amplitudes and periods for each single-shot realization. We do not expect that this~1D regime would be fully entered for most of the data in Fig.~\ref{fig:1000shots}
since that would require a condensate of about~500 atoms or less, even for the harmonic potential ($\omega_x=\rm2\pi\times45\,Hz$; $\omega_y\approx\omega_z\approx\rm2\pi\times950\,Hz$, Sec.~\ref{subsec:setup})~\cite{Gorlitz2001}. Nevertheless,~1D effects can be expected even above this limit for a sufficiently high temperature~\cite{Dettmer2001}. We were unable to determine the temperature experimentally (see Appendix~\ref{sec:appendix_contrast}). We consider~1D effects upon estimates of the correlation length in our experiment in Sec.~\ref{subsec:length}. 

In Fig.~\ref{fig:1000shots}(b) we compare the fringe contrast of the observed diffraction side-peaks to the contrast expected from simulated averages over a given number of experimental realizations, each with high contrast but with varying fringe positions due to random phases amongst the potential wells. It is apparent that the experimental data lie~2-5 standard deviations beyond the combined simulated and experimental errors, even for the longest trap holding times (the data of Fig.~\ref{fig:30shots}(b) correspond to an even higher number of standard deviations). We conclude that the signals observed in Fig.~\ref{fig:1000shots}(a) cannot arise from random phases for each experimental realization and must instead arise from spatial coherence over at least several magnetic lattice sites, as quantified further in Sec.~\ref{subsec:length}.  

\section{Analysis\label{sec:analysis}}

\subsection{Contrast\label{subsec:contrast}}

The observed fringe patterns are asymmetric and their three peaks appear to have different widths, making it difficult to fit them with simple functional forms. Low signal-to-noise ratios further preclude reliable fitting and unambiguous identification of the three peaks for single-shot images.  We therefore apply the following automated algorithm to analyze averages of at least~25 experimental images: (a)~the central peak is identified; (b)~secondary maxima are identified on the left and right sides if they occur with a separation of~6-20$\rm\,\mum$ from the central peak; (c)~if a secondary maximum cannot be found then the contrast on that side is defined to be zero; (d)~if there are two or more maxima on one side then the highest one is chosen; (e)~minima are identified between the side-peak maxima and the central peak. The contrast is then calculated by defining an~$\rm OD_{max}^{int}$ value which is interpolated between the central and side peaks at the position of the minimum on each side (\ie\ we define a triangular envelope on each side of the central peak). The contrast is then defined as the mean of~$\rm(OD_{max}^{int}-OD_{min})/(OD_{max}^{int}+OD_{min})$, again averaged over all the experimental images for a given holding time. Error bars are extracted from a bootstrapping procedure~\cite{Efron1986}.

An ideal analysis of the decoherence time in our experiment would require that all experimental parameters that could affect the contrast are independent of the holding time. In particular however, we cannot maintain the same number of atoms~$N$ due to atom loss, as seen in Fig.~\ref{fig:1000shots}(a). This loss can affect the observed contrast in several ways, including the fact that fewer atoms occupy less sites so that the coherence increases for a given coherence length, and that more atoms increase the density between sites, giving rise to Fourier components that diminish the contrast. Furthermore, one should not exclude the possibility that~$N$-dependent non-linear effects (\eg\ repulsion) are taking place during the launch/expansion time. 

To quantify the result of these effects, we count~$N$ for each of the individual images collected in Fig.~\ref{fig:1000shots}(a). We integrate the optical density over~$-60<x<60\rm\,\mum$ and~$-50<z<50\rm\,\mum$ and use the absorption cross section for~$\rm^{87}Rb$, after subtracting a sloping background that is linearly interpolated from data beyond these regions of significant atom density. Each image is then binned into successive ranges of~$N$, such that each bin for a given holding time uses the same number of images. Finally, the average contrast for the images in each bin is analyzed as described above, and the mean number of atoms~$\langle N\rangle$ is calculated for that bin.

\begin{figure}[t!]
   \centering
   \includegraphics[width=0.5\textwidth]{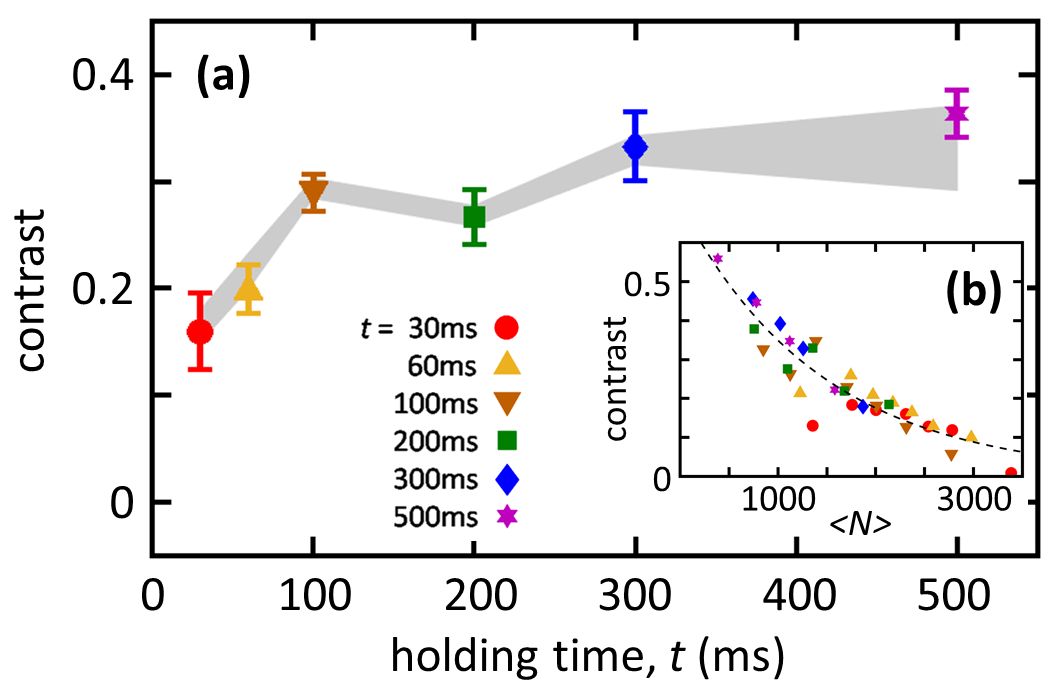}
   \caption{(Color online) Persistent spatial coherence. (a)~Average contrast of the two first-order fringes for all images containing~$1100<N<1900$ atoms. Error bars are extracted from a bootstrapping procedure~\cite{Efron1986} that yields the experimental statistical errors for each holding time. The shaded band shows systematic errors corresponding to several different ranges of~$N$ (see text). (b)~Average contrast of the two fringes, showing a systematic dependence on~$N$, hence requiring the restricted range used in~(a). See text for possible sources of this dependence. The dashed line is an exponential fit to guide the eye; error bars are omitted for clarity. Color code and symbol types as in Fig.~\ref{fig:1000shots}.} 
   \label{fig:contrast}
\end{figure}

The range of~$1100<N<1900$ atoms gives the best overlap between the highest~$N$ for~$t=\rm500\,ms$ and the lowest~$N$ for~$t=\rm30\,ms$ (\ie\ there are very few experimental images yielding~$N<1100$ atoms for~$t=\rm30\,ms$ or~$N>1900$ atoms for~$t=\rm500\,ms$). This range is sufficiently narrow to remove most of the~$N$ dependence, but wide enough that it includes~35-75 images for each holding time~$t$. We then plot the contrast as a function of the holding time in Fig.~\ref{fig:contrast}(a). The strong dependence of contrast on~$N$ that is evident from Fig.~\ref{fig:contrast}(b) clearly justifies this post-acquisition control of~$N$ for quantitatively comparing the observed contrast \vs\ holding time~$t$.

To verify the stability of this analysis, we re-calculated the contrast for several intervals of~$N$.  The shaded band in Fig.~\ref{fig:contrast}(a) includes results for the ranges~$N$=1200-1800, 1300-1900, 1100-1800, and~1350-1950, where these ranges were chosen such that the sample used would have at least~25 experimental cycles for each~$t$.  The shaded band of the figure shows that the observed behavior of the contrast does not depend on the range of~$N$ used. Further checks confirming the validity of this analysis are described in Appendix~\ref{sec:appendix_contrast}. We conclude that the spatial coherence is temporally robust, with no loss of contrast for at least~$\rm500\,ms$ (\ie\ no dephasing).

\subsection{Coherence length\label{subsec:length}}

\begin{figure}[t!]
   \centering
   \includegraphics[width=0.4\textwidth]{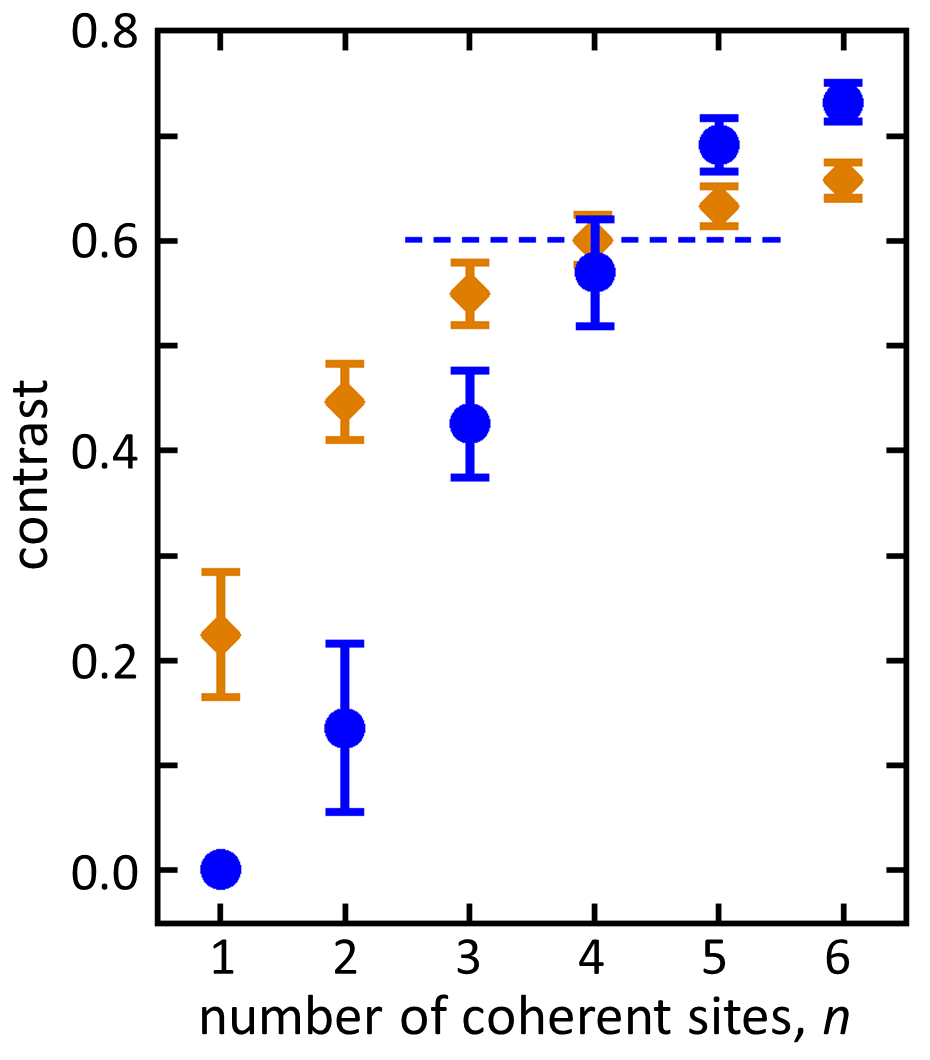}
   \caption{(Color online)
   Coherence length. Comparison of the observed contrast (dashed line) to two simple numerical models for the contrast expected when averaging~30 images. The~BEC is assumed to be coherent over a given number of lattice sites~$n$ and random for the remainder (blue points), or it is assumed that the phase correlation drops exponentially over a~$1/e$ distance of~$n$ sites (yellow points). In both cases, the observed contrast corresponds to a spatial coherence extending over~$\approx4$ sites, \ie~$\rm15\,\mum$.}  
   \label{fig:cohlength}
\end{figure}

In Fig.~\ref{fig:cohlength} we examine the spatial coherence length by comparing the observed contrast to that calculated from two simple numerical models. In our experiment the atoms are characterized essentially as a~3D condensate with potential barriers progressively higher relative to the chemical potential for magnetic lattice sites furthest from the center [Fig.~\ref{fig:schematic}(c)]. Consequently, the outermost sites could be fully disconnected, resulting in coherence only for the innermost lattice sites. In this case, the most relevant parameter for characterizing the coherence is roughly the number of coherently connected sites~$n$, where the phase is assumed to be constant over~$1,2,\dots,n$ central lattice sites and random over the remaining sites. A second model corresponds more directly to the notion of ``coherence length'' by accounting for possible cases in which the condensate is (at least partially) connected over the whole range but coherence is lost due to various factors. In this model, we assume partially random phases between the lattice sites such that the correlation between them drops exponentially over a~$1/e$ distance of~$n$ sites.

The results of both models are shown in Fig.~\ref{fig:cohlength}. The error bars represent statistical deviations and an additional uncertainty in the width and central position of the atomic distribution among the sites. For the first model, these error bars become smaller when the phase is completely random~($n=1$) or almost completely constant~($n=6$). Our optical resolution limits the maximal contrast for~$n=6$. 

The dashed line represents the contrast obtained from the experimental data of Fig.~\ref{fig:30shots}(b) (averaged over the left and right peaks), and corresponds to an observed spatial phase coherence over about~4 sites for both models. Since a reduction in the observed contrast may be caused by a variety of reasons as discussed above, we conclude from our data that the spatial coherence length is at least~$\rm15\,\mum$. Given the proximity of our trap to the surface, this observed coherence length is a significantly greater distance than expected if decoherence arises from, for example, Johnson noise (the correlation length of Johnson noise is expected to be about~$\rm5\,\mum$ in our experiment~\cite{Henkel2003}).

\section{Theory\label{sec:theory}}

\subsection{Simulation of atom dynamics\label{subsec:simulation}}

In order to gain a quantitative understanding of the experimental results, we have performed an extensive simulation of the dynamics of the atoms during the experiment, which takes into account the atom-atom interactions through the mean-field \GP~(GP) theory. Here we describe the simulation procedure and present some of its results. The simulation is intended to mimic the experimental conditions in a realistic manner, but it does not take into account possible systematic or random non-ideal effects due to fabrication defects or mis-alignment of system elements or imperfections in the preparation of the~BEC. Such imperfections may cause some of the asymmetry of the observed diffraction patterns, as well as some of the observed loss of contrast, but a detailed exploration of these effects is beyond the scope of the present paper.

The magnetic potential is calculated by applying the Biot-Savart law for the current density in the atom chip wires, as derived from a finite-element solution for the measured wire geometry. Unintended potential corrugations due to imperfections of the edge or bulk of the ``snake'' wire are not taken into account. The magnetic potential at a distance~${\bf r}$, a few~$\rm\mum$ from the snake wire, has the approximate form
\begin{eqnarray} V({\bf r}) &\approx& \frac{1}{2}\,m\left[\omega_x^2 x^2+\omega_y^2y^2+\omega_z^2(z-z_0)^2\right]
\nonumber \\ && +V_0\ e^{-(z-z_0)/l}\sin(kx), \label{eq:Vsnake} \end{eqnarray}
where~$\omega_x$ and~$\omega_y\approx\omega_z$ are the (unmodulated) longitudinal and transverse frequencies respectively, $z_0\approx\rm5\,\mum$ is the trapping distance, and~$k=2\pi/\lambda$ gives the periodicity of the potential modulation with~$\lambda=\rm5\,\mum$. The amplitude of the modulation decreases exponentially with the atom-surface distance, with a range parameter of~$l\approx\rm1\,\mum$; in addition,~$V_0$ and the transverse frequencies increase parametrically as the trapping distance~$z_0$ decreases. Modulation of the longitudinal potential gives rise to a series of traps as shown in Fig.~\ref{fig:schematic}(c), with each site having a frequency of~$\omega_{\rm site}\approx k\sqrt{V_0/m}\sim2\pi\times\rm500\,Hz$, about half of the transverse frequencies~$\omega_y$ and~$\omega_z$.

\begin{figure}[t!]
   \centering
   \includegraphics[width=0.50\textwidth]{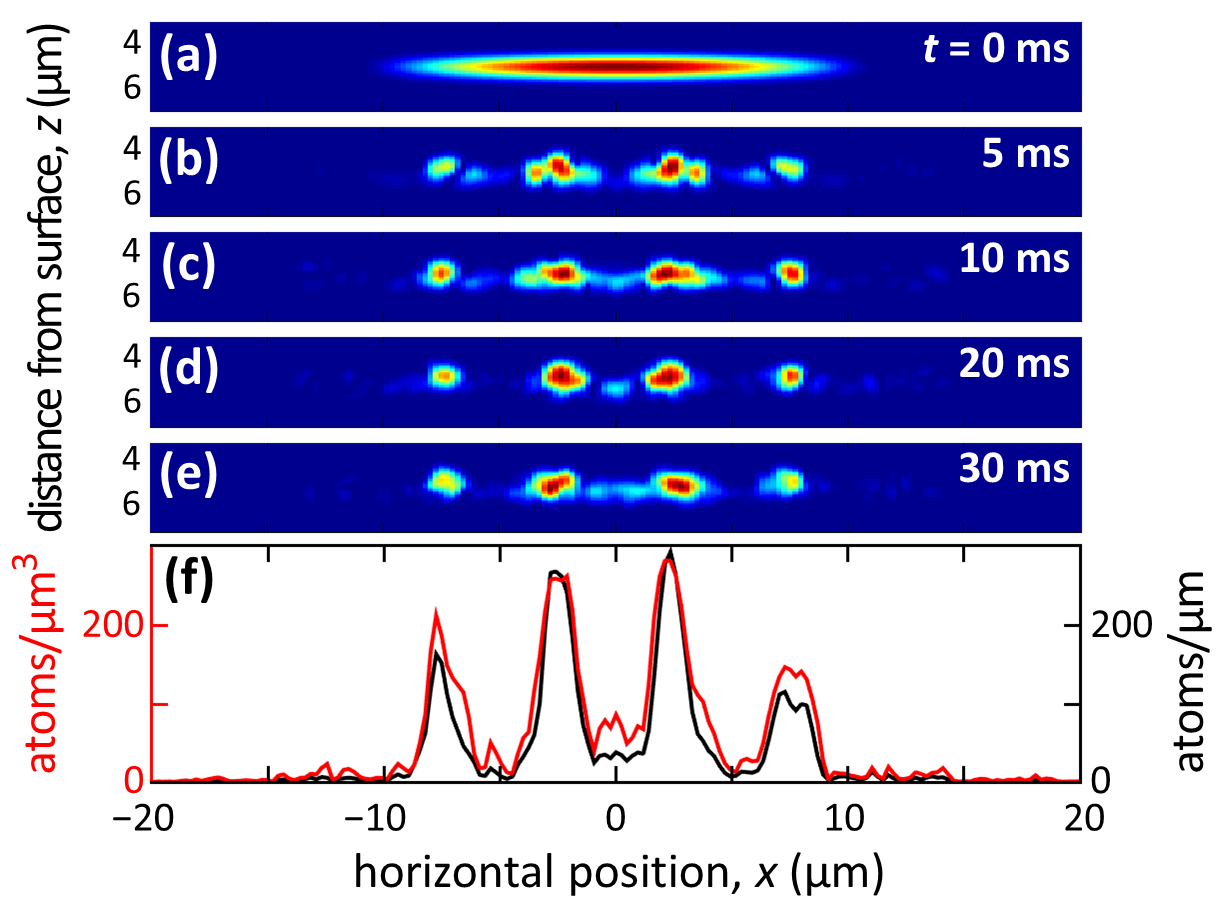}
   \caption{(Color online) Simulated evolution during the holding time. (a)~Initial ground-state~BEC with~$N=1500$ atoms in the nearly harmonic (unmodulated) potential. (b-e)~Re-distribution of the atomic density as a function of the holding time~$t$ in the magnetic lattice, shown as atomic column densities in the~$x$-$z$ plane. (f)~Atomic density per unit volume (red) and unit length (black) are shown as an average over~$t=\rm5,10,\dots,30\,ms$. In this simulation, the chemical potential slightly exceeds the barrier height (Sec.~\ref{sec:discussion}). The density between the barriers becomes smaller with horizontal positions further from the center due to the rising longitudinal harmonic potential.}
   \label{fig:holding}
\end{figure}

The simulation starts with~$N$ atoms in a~BEC ground state that mimics the atomic cloud before it is loaded into the modulated potential. The modulation is then ramped up in~$\rm1\,ms$ until the full potential is attained, at which time it may be approximated by Eq.~(\ref{eq:Vsnake}) with~$V_0=\rm40\,nK$ ($\rm80\,nK$ peak-to-valley). This ramp-up of the modulation provides an adequate approximation of the loading procedure, achieved experimentally by lowering the snake-wire current and thereby bringing the potential minimum to~$z_0$. After the potential modulation is turned on at time~$t=0$, the evolution in the trap is calculated up to~$\rm30\,ms$, corresponding to the shortest holding time used in our experiments. Snapshots of the atomic density in the trap for several times during this period are presented in Fig.~\ref{fig:holding}(a-e).

\begin{figure}[t!]
   \centering
   \includegraphics[width=0.45\textwidth]{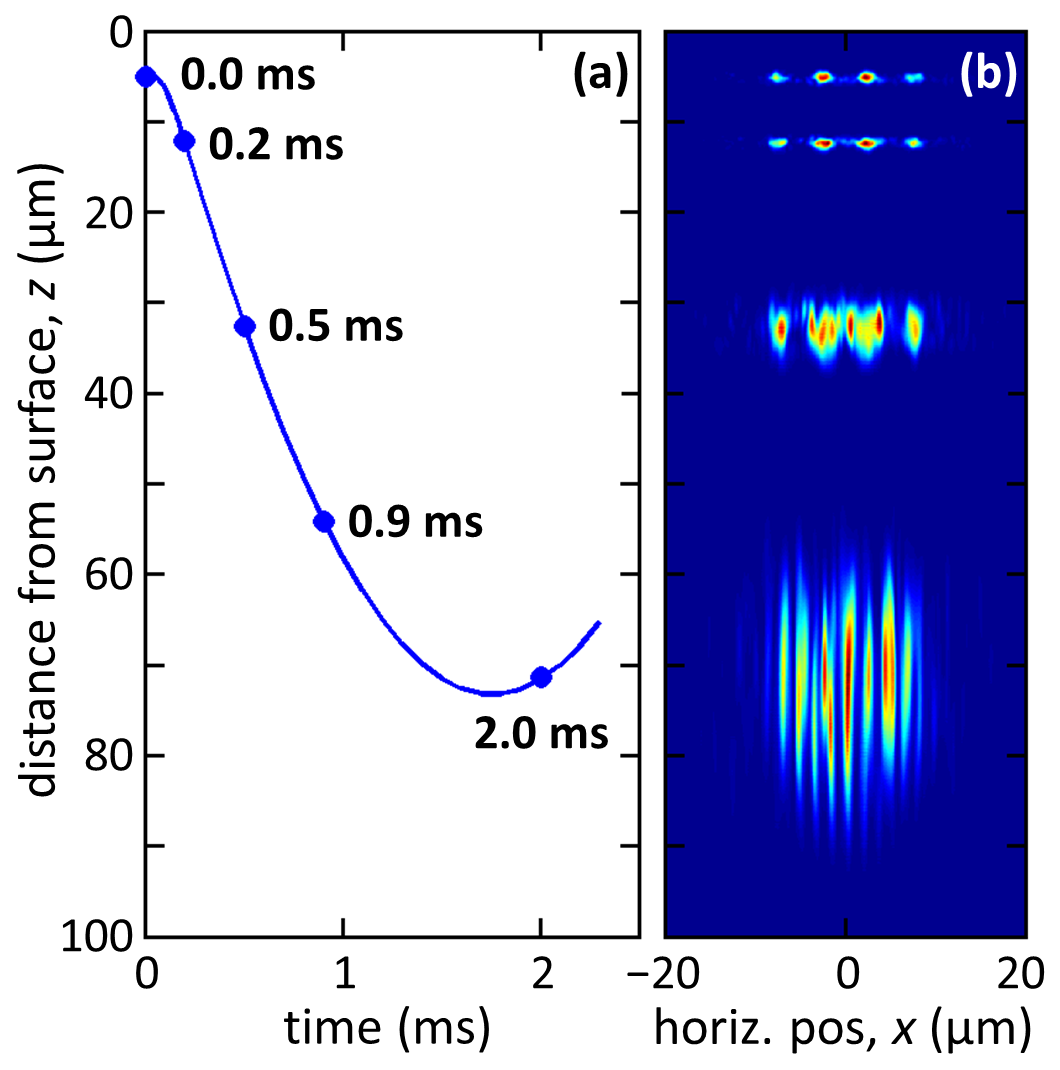}
   \caption{(Color online) Simulated evolution during the launch. The cloud is ejected from the magnetic lattice by increasing the current in the snake wire from~$\rm5.5\,mA$ to~$\rm18\,mA$ in~$\rm100\,\mus$ (Sec.~\ref{subsec:setup}). (a)~Trajectory of the center-of-mass of the cloud during the launch. The cloud arrives at the outer turning point (created by the trapping wire which is still on) and starts to accelerate back towards the atom chip before all currents are turned off (see text). Dots along the trajectory depict times at which the atomic density is shown in~(b). Full release occurs after~$\rm2.3\,ms$. (b)~The near-field diffraction pattern is seen to start forming already during the launching stage, while the atomic cloud starts to expand rapidly along the radial direction. Our optical resolution is currently unable to resolve detailed features of the expanding cloud during this period (see Fig.~\ref{fig:simulation} for the corresponding far-field simulation results.)}
   \label{fig:launch}
\end{figure}

After the holding time we simulate launching of the~BEC by ramping the current in the snake wire from~$\rm5.5\,mA$ to~$\rm18\,mA$ as described in Sec.~\ref{subsec:setup}. We solve the~GP equation in a frame of reference that moves together with the center-of-mass of the atomic cloud. The shape and position of the atomic cloud during this launching process is shown in Fig.~\ref{fig:launch}. After~$\rm2\,ms$ the cloud arrives at the outer turning point of the potential and starts to accelerate back up towards the atom chip. Another~$\rm0.3\,ms$ before the final trap release allows a longer period of time for the focusing and also increases the free-fall time before the cloud leaves the field-of-view of our imaging system. The simulated currents and magnetic fields are then turned off, as in the experiment, and the cloud is allowed to start falling freely in gravity.

\begin{figure}[t!]
   \centering
   \includegraphics[width=0.4\textwidth]{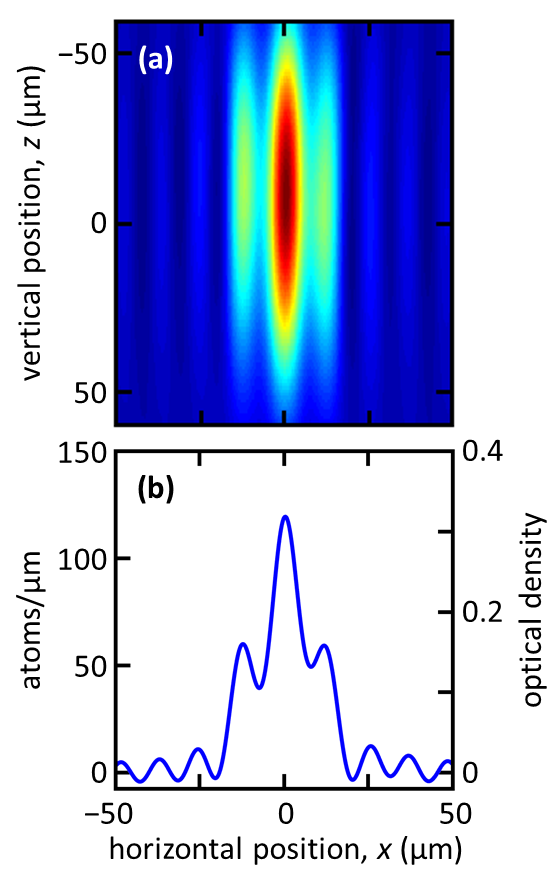}
   \caption{(Color online) Simulated diffraction pattern after release. (a)~The shape of the atomic cloud after~$\rm12\,ms$ of free fall in the~$x$-$z$ plane, and (b)~a cut through the center of the cloud along the horizontal direction~$x$. The finite resolution of an imaging system similar to ours is included (Appendix~\ref{sec:appendix_calib}). The observed asymmetry between the first-order peaks can arise from slight asymmetries in the magnetic potential near the snake wire due to the well-known rotation of the cloud by the Z-shaped trapping wire (Sec.~\ref{subsec:asymmetry}). Since this simulation was done with the full magnetic potential, the latter rotation is introduced automatically.}
   \label{fig:simulation}
\end{figure}

After releasing the cloud, we calculate its free expansion for a time-of-flight of~$\rm12\,ms$. The simulated cloud develops a diffraction pattern with a central (zero-order) peak and several diffraction orders (Fig.~\ref{fig:simulation}), where each diffraction order is squeezed into a narrow wavepacket in the~$x$-direction due to the focusing effect (described next). Here we show only the column density of the atoms as predicted by applying a low-pass filter to simulate the effect of our finite optical resolution (Appendix~\ref{sec:appendix_calib}). This far-field image can be compared to our experimental results, unlike the near-field simulated images of Fig.~\ref{fig:launch} which would not be experimentally resolvable.

\subsection{Focusing effect\label{subsec:focusing}}

\begin{figure}[t!]
   \centering
   \includegraphics[width=0.45\textwidth]{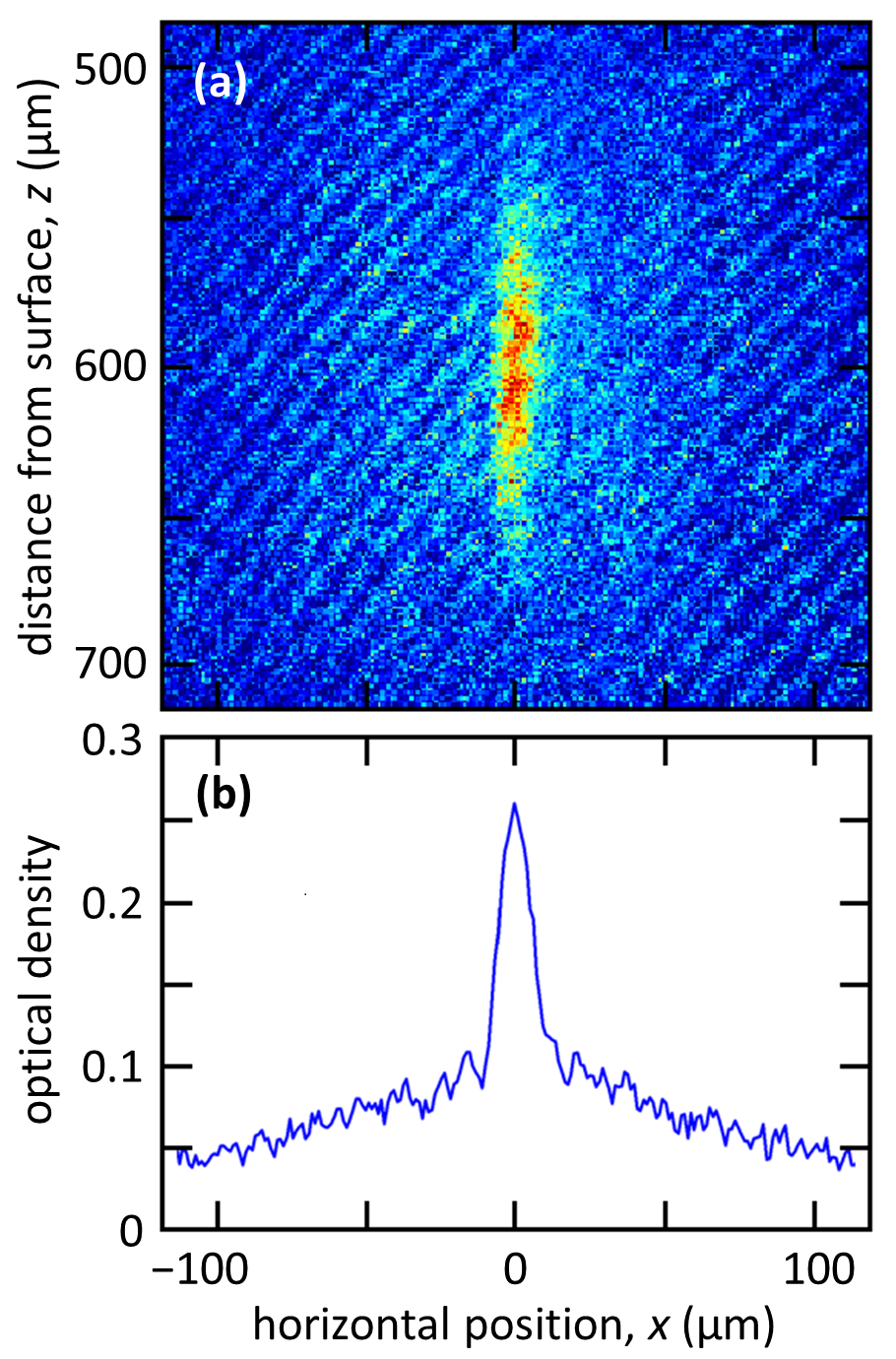}
   \caption{(Color online) Focusing effect. (a)~Single-shot image of a~BEC, launched and released under the same conditions used throughout this work, and acquired after~$\rm12\,ms$ of free-fall (time-of-flight). The~BEC is initially held in a trap~$\rm9\,\mum$ from the surface, where there is no potential modulation, and hence no fringes are seen in the image. (b)~A cut through the center of~(a) with a measured Thomas-Fermi full-width of~$18\,\mum$, compared to a width of about~$\rm35\,\mum$ calculated with no focusing. The thermal tails are seen to be unfocused.}
   \label{fig:focus}
\end{figure}

Figure~\ref{fig:focus} illustrates the focusing effect we observe for a~BEC initially held~$\rm9\,\mum$ from the surface. At this distance, the~$\rm5\,\mum$-period modulation potential has no influence [Fig.~\ref{fig:schematic}(c)], and the focusing effect may be understood by considering only the harmonic trapping potential. Under these circumstances, we measure a Thomas-Fermi full-width of~$18\,\mum$, considerably narrower than a width of about~$\rm35\,\mum$ expected in the absence of focusing.

Next we briefly explain the effect of focusing and its importance for the observation of diffraction patterns emerging from a~BEC that is density- or phase-modulated. First we consider the evolution of a general atomic wave packet in free space, which applies to our~BEC after full release, whereupon the atom-atom interactions are negligible due to rapid expansion of the atomic cloud in the transverse directions. In this case, the free-space evolution of the initial wave function~$\psi({\bf r},t_{\rm TOF}=0)$, in a reference frame moving with the freely falling center-of-mass, is given by
\begin{equation} \psi({\bf r},t_{\rm TOF})=\int d^3{\bf k}\, \exp\left[i\left({\bf k}\cdot{\bf r}-\frac{\hbar k^2 t_{\rm TOF}}{2m}\right)\right]
\tilde{\psi}({\bf k}),
\label{eq:psir} \end{equation}
\vskip0.5\baselineskip\noindent where\vskip-3.3\baselineskip
\begin{equation}
\tilde{\psi}({\bf k})=\frac{1}{(2\pi)^3}\int d^3{\bf r}'\, e^{-i{\bf k}\cdot{\bf r}'}\psi({\bf r}',0)
\label{eq:psik} \end{equation}
is the spatial Fourier transform of the initial wave function.

After a sufficiently long time, the expansion leads to a separation of the different momentum components; the far-field limit of the wave function is given by the stationary phase approximation, which yields
\begin{equation} \psi({\bf r},t_{\rm TOF})\propto e^{imr^2/2\hbar t_{\rm TOF}}\ \tilde{\psi}({\bf k}=m{\bf r}/\hbar t_{\rm TOF}),
\label{eq:fourier} \end{equation}
such that the atomic density represents the Fourier transform of the initial wave function.

More explicitly, by substituting Eq.~(\ref{eq:psik}) into Eq.~(\ref{eq:psir}) and integrating over the momentum~${\bf k}$, we obtain
\begin{equation}
\psi({\bf r},t_{\rm TOF})=\int d^3{\bf r}' G({\bf r}-{\bf r}',t_{\rm TOF})\psi({\bf r}',0), \end{equation}
\vskip0.0\baselineskip\noindent where\vskip-3.5\baselineskip
\begin{equation}\qquad\quad
G({\bf r}-{\bf r}',t)=\left(\frac{m}{2\pi i\hbar t}\right)^{3/2}\exp\left(im\frac{|{\bf r}-{\bf r}'|^2}{2\hbar t}\right)
\end{equation}
is the non-relativistic free-particle Feynman propagator. By expanding the square term in the exponent, we obtain \vskip-2\baselineskip
\begin{eqnarray}
\psi({\bf r},t_{\rm TOF})&=& \left(\frac{m}{2\pi i\hbar t_{\rm TOF}}\right)^{3/2}
e^{imr^2/2\hbar t_{\rm TOF}} \times
\nonumber \\ && \hspace{-75pt}				
\int d^3{\bf r}'\	
\exp\left(-i\frac{m}{\hbar t_{\rm TOF}}{\bf r}\cdot{\bf r}'\right)
\exp\left(i\frac{mr'^2}{2\hbar t_{\rm TOF}}\right)
\psi({\bf r}',0). \label{eq:integral} \end{eqnarray}
The second exponent of the integral in Eq.~(\ref{eq:integral}) represents a quadratic phase which decreases with time as~$1/t_{\rm TOF}$. To attain the far-field form of a Fourier transform of the initial wave function as in Eq.~(\ref{eq:fourier}) would therefore require a time~$t_{\rm TOF}\gg m\Delta x^2/2\hbar$, where~$\Delta x$ is the spatial extent of the initial atomic density along the~$x$ (longitudinal) direction. In our case, where~$\Delta x\approx\pm\rm15\,\mum$, this implies that a fully developed diffraction pattern would require~$t_{\rm TOF}>\rm150\,ms$. This is much too long to allow observation in our system.

In order to overcome this limitation, we implement a launching procedure, in which the atoms are kept in the harmonic potential for a time~$\tau=\rm2.3\,ms$ even while being pushed away from the atom chip. Whereas before the launching procedure the atomic density along the~$x$ direction is determined by the equilibrium between the confining modulated harmonic potential and the repulsive force of the atom-atom interactions, after the beginning of launching the~BEC expands in the radial direction and the repulsive force weakens rapidly. At this stage the harmonic force, no longer compensated by the strong repulsive potential, induces a velocity gradient along~$x$, such that the atoms start to move towards the center. For a short time~$\tau$ this velocity gradient is not sufficient to  change the atomic density significantly and the main effect is to imprint a quadratic phase $\phi=-\alpha x^2$,\negskip
\begin{equation} \psi({\bf r}',0)\to \psi({\bf r}',0)e^{-i\alpha x^2}, \end{equation}
where $\alpha=\frac{1}{2}m\omega_x^2\tau/\hbar$.
When the atoms are finally released, this quadratic phase partially compensates for the quadratic phase in the second exponent of the integral in Eq.~(\ref{eq:integral}), such that after a finite time-of-flight~$t_{\rm TOF}=m/2\hbar\alpha\sim(\omega_x^2\tau)^{-1}$ the total quadratic phase in the integral vanishes completely, leading to a density pattern along the~$x$ direction that represents the Fourier transform of the initial density before launching, \ie\ a fully developed diffraction pattern as expected in the far-field limit.

The effect of the harmonic potential during the launching stage on matter waves is equivalent to the effect of a focusing lens on an incident optical beam, namely, focusing an incident plane wave and Fourier transforming an arbitrary input at the focal plane. If the initial atomic cloud is a coherent smooth~BEC with a narrow momentum distribution then the process leads to a focused cloud as in Fig.~\ref{fig:focus}. In the case of an initial cloud with a modulated density as in the main part of this paper, we obtain a series of focused diffraction peaks as in Fig.~\ref{fig:30shots}. In contrast, the incoherent thermal part of the initial cloud, which has a wide initial momentum distribution (emulating an optical beam with random~$k$ vectors impinging on the lens), is not focused but rather continues to expand after the launch and release, shown as the~$\sim\rm100\,\mum$-wide background distribution in Fig.~\ref{fig:focus}.

Finally, the time-of-flight actually required for focusing is somewhat longer than the value of~$(\omega_x^2\tau)^{-1}\approx\rm5.4\,ms$ noted above because repulsive forces do play a limited role during the initial part of the evolution, \ie\ in the launching. During this time, the mechanism of the focusing process is affected by interatomic interactions within the~BEC and therefore differs from atomic lensing for non-interacting atoms~\cite{Smith2008,Zhou2009}. Our~GP simulation described in Sec.~\ref{subsec:simulation} predicts that the best focusing should occur instead at about~$t_{\rm TOF}=\rm8\,ms$. We actually use a time-of-flight of~$t_{\rm TOF}=\rm12\,ms$ in order to allow further separation of the diffraction fringes; this additional delay does not significantly affect the sharpness of the diffraction peaks we observe, which is limited anyway by our finite optical resolution.

\subsection{Asymmetry due to phase gradients and randomness \label{subsec:asymmetry}}

\begin{figure}[t!]
   \centering
   \includegraphics[width=0.45\textwidth]{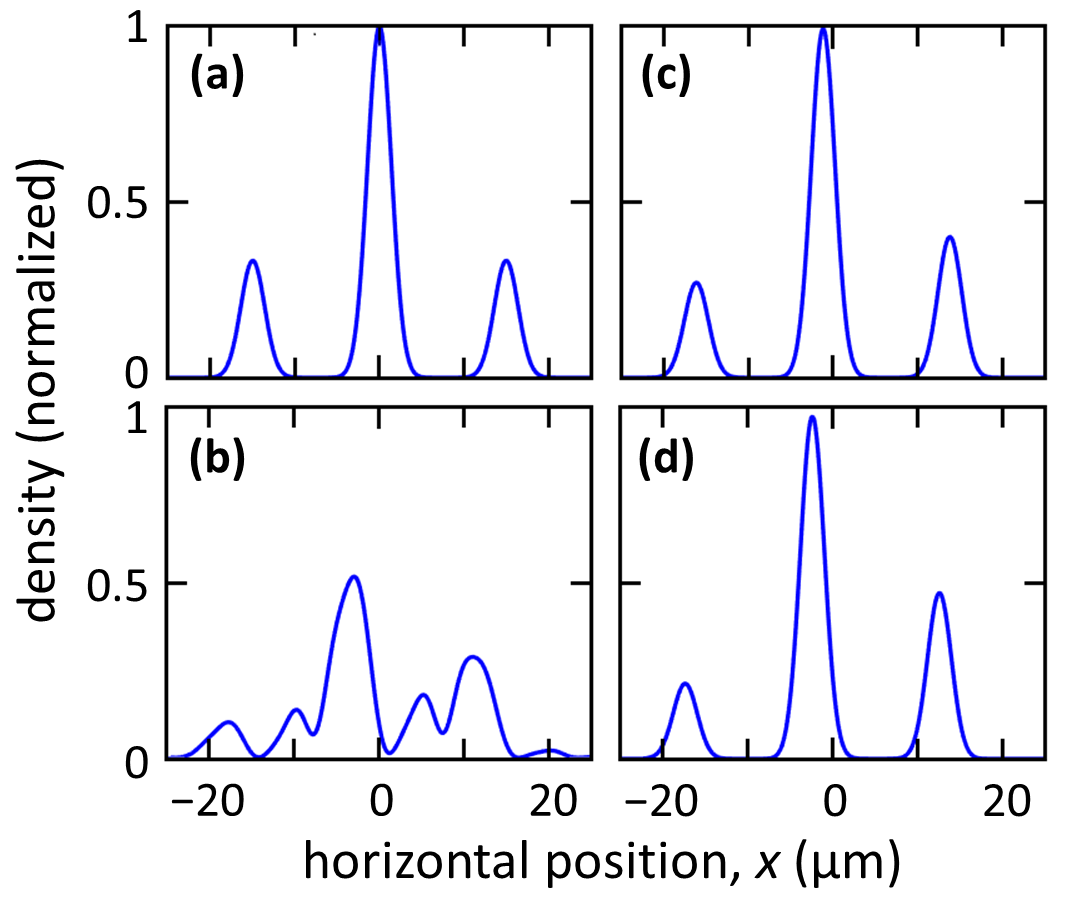}
   \caption{(Color online) A simple model for the formation of an asymmetric diffraction pattern. (a)~A symmetric diffraction pattern with two first-order peaks is formed by a Fourier transform of a wave function with a periodic amplitude and a constant phase. (b)~A noisy diffraction pattern is obtained by applying a random phase between the sites. (c-d)~Asymmetry between the two first-order peaks is obtained when a linear phase difference is applied between the sites (while keeping a constant phase within each site). Here~(c) and~(d) are obtained from a phase difference of~$\rm0.5\,rad$ and~$\rm1\,rad$ between adjacent sites, respectively (see text for possible sources of phase differences between sites). Note that, in addition to asymmetry, the fringe patterns in~(c) and~(d) also shift horizontally relative to~(a).}
   \label{fig:asymm}
\end{figure}

We study the effect of phase randomization and phase gradients over the wave function in the modulated potential by using a simple numerical model. Quasi-condensate phase fluctuations, characteristic of elongated traps~\cite{Dettmer2001}, are not considered since our model greatly exaggerates these effects for clarity. We start with a one-dimensional density function~$\rho(x)$ composed of a Gaussian function centered at~$x=0$ and two side-band Gaussians at~$x=\rm\pm15\,\mum$ whose maxima are one-third that of the main peak, as shown in Fig.~\ref{fig:asymm}(a). This density distribution represents a diffraction pattern that is produced by a wave function with a periodically modulated amplitude and a constant phase over all the sites of the magnetic lattice. We introduce phase changes between the sites by first Fourier transforming the square root of the triple-Gaussian function~$\sqrt{\rho}$ to obtain the periodically modulated function~${\cal F}\{\!\sqrt{\rho}\}$ and then introducing abrupt phase changes at the minima of this function, where the density is negligible. The phase is therefore constant within each site and varies only between sites. An inverse Fourier transform then re-generates the density function~$\rho\,'(x)$ as modified by the phase changes imposed by our numerical model.

By introducing random phases between sites, we obtain noisy diffraction patterns like the one presented in Fig.~\ref{fig:asymm}(b). Note that it becomes difficult to identify the zero-order and first-order peaks unambiguously and that smearing reduces the~OD of the central peak by a factor of~$\approx2$ even for individual patterns. Averaging a given number of similarly noisy patterns, each with its own distribution of phases between sites, is used to generate the theoretical line and band shown in Fig.~\ref{fig:1000shots}(b) and the data points shown in Fig.~\ref{fig:cohlength}. 

By introducing a linear phase gradient instead, \ie\ equal phase jumps between the sites, we obtain the asymmetric diffraction patterns shown in Fig.~\ref{fig:asymm}(c-d). This may explain some of the asymmetry observed in our experimental results. Other contributions to the observed asymmetry may include a slight rotation of the atomic cloud by the trapping wire (a well-known effect produced by currents in Z-shaped wires, as seen in
Fig.~\ref{fig:simulation}), and possible imperfections in the snake-wire fabrication~\cite{Preston1970}.

\section{Discussion\label{sec:discussion}}

We now examine in more detail the dephasing taking place in our system. Figure~\ref{fig:contrast} shows the fringe contrast as a function of holding time~$t$, after accounting for the observed dependence between the contrast and the atom number~$N$ (Sec.~\ref{subsec:contrast}). This provides a direct estimate of the relative coherence for the different holding times, assuming that experimental imperfections reducing the contrast are independent of the holding time.  

The data of Fig.~\ref{fig:contrast}(a) are consistent with little or no dephasing. The apparent slight rise of contrast with time may have several origins. For example,~BEC excitations due to imperfect loading may be relaxing. In addition, thermal atoms may be escaping to the surface \via\ evaporation due to the weak trap depth of only~$\approx2\,\muK$~\cite{Harber2003}, thus increasing the proportion of the atomic cloud that is coherent. Losing thermal atoms may also decrease phase fluctuations of the~BEC~\cite{Dettmer2001}. These processes are consistent with the fact that the strongest rise in contrast is observed in the first~$\rm100\,ms$. Further analysis of this slight rise is given in Appendix~\ref{sec:appendix_contrast} but an exact determination of its origins is left for future work.

A crucial parameter for understanding dephasing is the chemical potential relative to the magnetic potential barrier height in our system. We estimate that~$V_{\rm barrier}\approx\rm80\,nK$ at a distance of~$\rm5\,\mum$ from the chip, which is very close to the longitudinal chemical potential (the single-atom effective energy, excluding the transverse energy), estimated from mean-field~(GP) calculations to be~$\mu_\parallel=\rm88\,nK$ for~$N=1500$ atoms. These estimates presume that the central barrier is located at the minimum of the harmonic potential; they both increase by~$\lesssim\rm5\,nK$ if the central barrier is shifted by up to half the lattice period. For the atoms sampled in Fig.~\ref{fig:contrast}(a), these estimates suggest that the~2-3 central wells are probably classically connected (Fig.~\ref{fig:holding} shows that the atom density above the barriers falls by a factor of~$\approx5$), while adjacent wells might be completely classically disconnected. The~GP calculations also show that the central wells have approximately double the population of the two adjacent wells.
 
\begin{figure}[t!]
   \centering
   \includegraphics[width=0.5\textwidth]{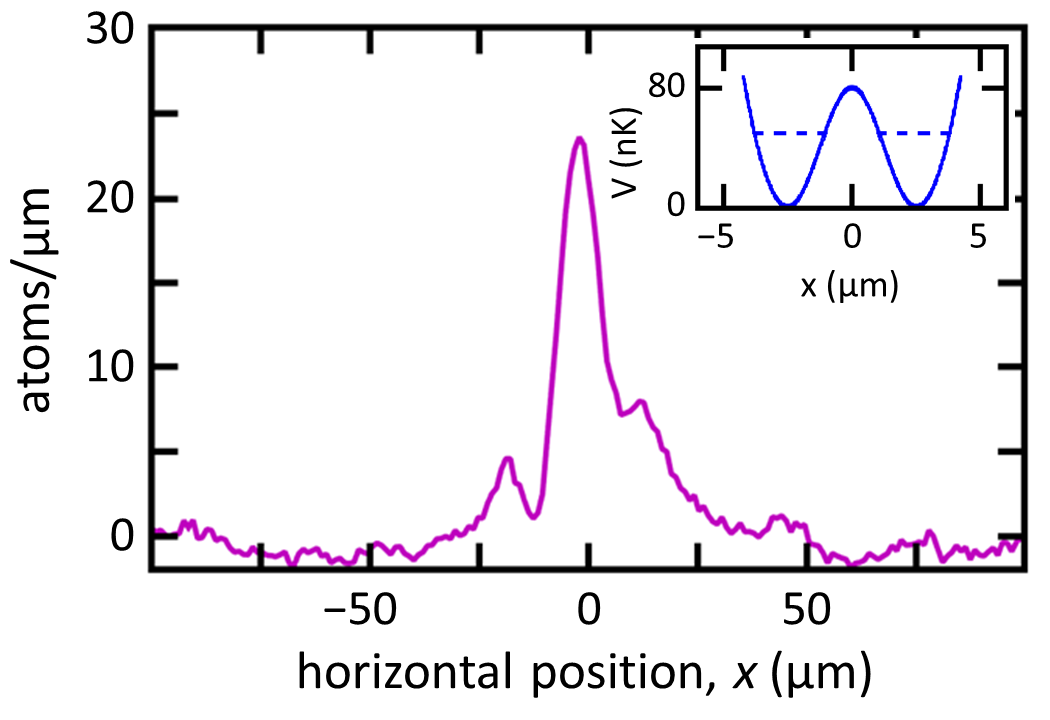}
   \caption{(Color online) Spatial coherence for low chemical potential. Diffraction pattern observed after a trap holding time of~$t=\rm500\,ms$ for an average of all~29 experimental cycles having~$N=400\pm130$ atoms, corresponding to the smallest number of atoms in Fig.~\ref{fig:contrast}(b). Note that the chemical potential is most likely below the central potential barrier, as depicted in the inset (dashed line, see text).}
   \label{fig:lowN}
\end{figure}

As a contrasting case, in which the chemical potential may be significantly below the potential barrier, we plot in Fig.~\ref{fig:lowN} an average of all~29 images having a holding time~$t=\rm500\,ms$ with the smallest number of atoms,~$N=400\pm130$. This range of~$N$ corresponds to a chemical potential of~$\mu_\parallel=\rm50\pm5\,nK$. The observed high contrast implies either that the chemical potential exceeds~$V_{\rm barrier}$, or that spatial coherence is maintained even if~$\mu_\parallel<V_{\rm barrier}$, as depicted in the inset. We note that our measurements of the distance to the chip are uncertain by about~$\rm\pm0.5\,\mum$ (Sec.~\ref{subsec:setup}), corresponding to an uncertainty of about~$\rm\pm40\,nK$ in~$V_{\rm barrier}$. Theory predicts that even if the chemical potential is below the barrier height, coherence should still be maintained. For example, calculations based on a simple double-well model show that a~BEC of~250-300 atoms would be in the Josephson interaction regime for~$V_{\rm barrier}\approx\rm60\,nK$ (well within the range of our experimental uncertainty), assuming that the~BEC is in the ground state~\cite{Japha2016}. In this regime the tunneling rate is sufficiently fast to maintain coherence, even though the chemical potential is significantly below the barrier. In addition, a higher barrier implying full~BEC separation at~$t=\rm500\,ms$ may still allow coherence to be maintained for a long time due, for example, to number squeezing~\cite{Jo2007,Berrada2013}.

It is worth noting that the~1D regime would be reached if fewer than~500 atoms were spread along the entire length of the trap, thereby reducing the observed contrast as discussed in Sec.~\ref{sec:results}. Figure~\ref{fig:lowN} shows no such reduction, suggesting instead that indeed the chemical potential is below the barrier, where the tighter confinement within individual lattice sites prevents quasi-condensate phase fluctuations while sufficiently fast tunneling maintains the observed phase coherence~\cite{Japha2016}.

\section{Summary and conclusions\label{sec:summary}} 

In conclusion, by loading a~BEC into a lattice potential~$\rm5\,\mum$ from a room temperature surface, followed by its careful release, we have demonstrated diffraction, a hallmark of spatial coherence, for an atom-surface distance reduced by an order of magnitude from previous experiments exhibiting diffraction or interference~\cite{Wang2005,Schumm2005a,Jo2007,Baumgartner2010}. In addition, our data exhibit robust spatial coherence that persists for a relatively long time,~$\tau_{\rm coh}\ge\rm500\,ms$, with a coherence length that is a significantly greater distance than the atom-surface separation. While spin coherence close to a room temperature environment has been demonstrated in dilute gases~\cite{Treutlein2004} and even in solid-state systems~\cite{Farfurnik2015}, maintaining spatial coherence only~$\rm5\,\mum$ from the surface constitutes another significant milestone for atom chip applications and for interferometric probing of surface effects, including unique features such as correlation lengths of forces and their fluctuations~(\cite{Keil2016x} and references therein).

These results should motivate further investigations to elucidate the interplay amongst tunneling, atomic collisions, and effects due to external noise, and their role in maintaining robust spatial coherence~\cite{Henkel2004,Boukobza2009,Khodorkovsky2009,Japha2016}. This experiment also holds promise for the future development of atomic circuits. Indeed, as shown quantitatively in previous work~\cite{Salem2010}, good control over tunneling barriers is possible for distances of~$\rm5\,\mum$ and below. In future it may be beneficial to use nanowires in order to considerably increase the magnetic field gradients so that they can overcome the Casimir-Polder potential at smaller distances, or to utilize molecular conductors such as carbon nanotubes~\cite{Petrov2009} or graphene sheets~\cite{Judd2011}, both of which would reduce Johnson noise and potential corrugations due to electron scattering.

\begin{acknowledgments}
We thank Julien Chab\'e, Ran Salem, and Tal David for the initial steps of the experiment, Zina Binstock for the electronics, and the~BGU nano-fabrication facility for providing the high-quality chip. We are grateful to Amichay Vardi and Carsten Henkel for helpful discussions. This work is funded in part by the Israel Science Foundation (1381/13), the European Commission ``MatterWave'' consortium (FP7-ICT-601180), and the German-Israeli DIP project (FO~703/2-1) supported by the Deutsche Forschungsgemeinschaft. We also acknowledge support from the~PBC program for outstanding postdoctoral researchers of the Council for Higher Education and from the Ministry of Immigrant Absorption (Israel).\negskip
\end{acknowledgments}

\appendix

\section{Atom-surface distance calibration and imaging resolution\label{sec:appendix_calib}} 

Our imaging resolution of about~$\rm7\,\mum$ is insufficient to resolve the \insitu\ atomic density modulation along the~$x$-axis expected from the~$\rm5\,\mum$-spacing between individual magnetic lattice sites. The imaging resolution is however, sufficient to resolve the atom cloud and its reflection along the~$z$-axis, produced by grazing incidence of the imaging laser beam on the reflective atom chip surface. Since the expected separation of these images~($\rm10\,\mum$) is only slightly greater than our imaging resolution, we must estimate the small systematic difference between the apparent distance obtained from the \insitu\ images and the actual distance of the atom cloud from the surface. 

This systematic difference is estimated by using \GP\ calculations to simulate a~BEC of~4000 atoms centered~$\rm4.8\,\mum$ from the atom chip reflective surface. We apply a low-pass filter for the spatial spectrum of the absorption pattern for direct comparison to the experimentally-measured optical density~(OD): \vskip-1.5\baselineskip
\begin{eqnarray}
{\rm OD}(x,z)&=& -2\log\left[\left|{\cal F}^{-1}\left\{{\cal F}\left\{e^{-n(x,z)\sigma_0/2}\right\}\times
\right.\right.\right. \nonumber \\ && \left.\left.\left.		
\theta(k_0^2-k_x^2-k_z^2)\right\}\right|\right],
\end{eqnarray}\vskip-0.5\baselineskip
\noindent where~$\sigma_0=\rm0.19\,\mum^2$ is the scattering cross-section of light with~$\rm^{87}Rb$ atoms,~$n(x,z)$ is the
column density of atoms,~${\cal F}$ and~${\cal F}^{-1}$ represent the Fourier and inverse Fourier transforms, and~$\theta(k_0^2-k^2)$ is the Heaviside function which allows only spectral components with wave vector smaller than~$k_0$ to pass through the system due to the finite aperture of the imaging lens. We estimate that $k_0\approx\rm2\pi/11.5\,\mum$. This procedure is equivalent to a convolution with an Airy function of about~$\rm7\,\mum$ radius representing the finite aperture of a lens (our diffraction-limited resolution is about~$\rm4\,\mum$ but this limit is not attained due to optical aberrations and shadowing of the lens by the atom chip). 

\begin{figure}[t!]
   \centering
   \includegraphics[width=0.45\textwidth]{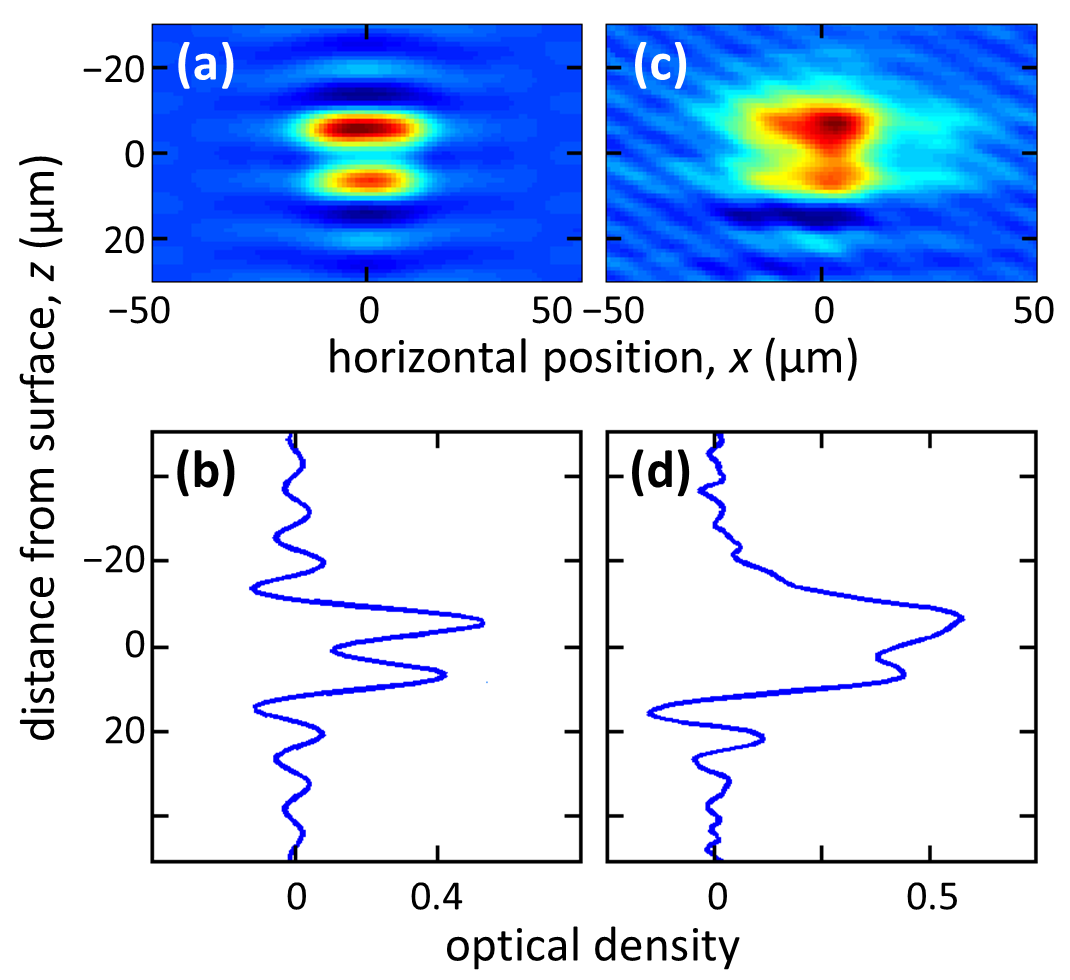}
   \caption{(Color online) Atom-surface distance measurements: imaging the atomic cloud in the trap near the chip. (a)~This simulated image shows the cloud and its reflection from the chip surface; a vertical cut through the center of the cloud is shown in~(b). (c-d)~The corresponding cloud and vertical cut from experimental measurements after a holding time of~$t=\rm30\,ms$ in the modulated potential. The detection efficiency for the reflected cloud image is assumed to be reduced by~30\% relative to the direct image. The peak positions of the simulation are chosen to match those of the experimental measurements and show a systematic difference between the measured atom-surface distance and the actual distance of about~$\rm0.8\,\mum$ (see text).}
   \label{fig:distance}
\end{figure}

The simulated results are shown in Fig.~\ref{fig:distance}(a-b), demonstrating partial resolution of the image and its reflection, as well as optical interference fringes due to a separation comparable to the cutoff wavelength of the optical system. The equivalent experimental absorption images are shown in Fig.~\ref{fig:distance}(c-d) after a holding time of~$t=\rm30\,ms$ in the modulated potential. The image is smoothed in Fig.~\ref{fig:distance}(d) with a Gaussian kernel~5 pixels wide in the horizontal~($x$) direction in order to reduce the experimental noise level. Double-Gaussian fits to the curves in Fig.~\ref{fig:distance} yield a separation of~$\rm11.3\,\mum$ for both the simulated and experimental images, corresponding to an apparent atom-surface distance of~$\rm5.6\,\mum$. We conclude that there is a systematic difference between the measured and actual atom-surface distance, with the actual distance being about~$\rm0.8\,\mum$ less than the measured distance. Atom-surface distances quoted in this paper are calibrated by this amount. We note that maintaining this distance within tight bounds is necessary for accurately determining the modulation of the magnetic lattice potential [Fig.~\ref{fig:schematic}(c)].

\section{Dependence of contrast on atom number\label{sec:appendix_contrast}}

\begin{figure}[t!]
   \centering
   \includegraphics[width=0.45\textwidth]{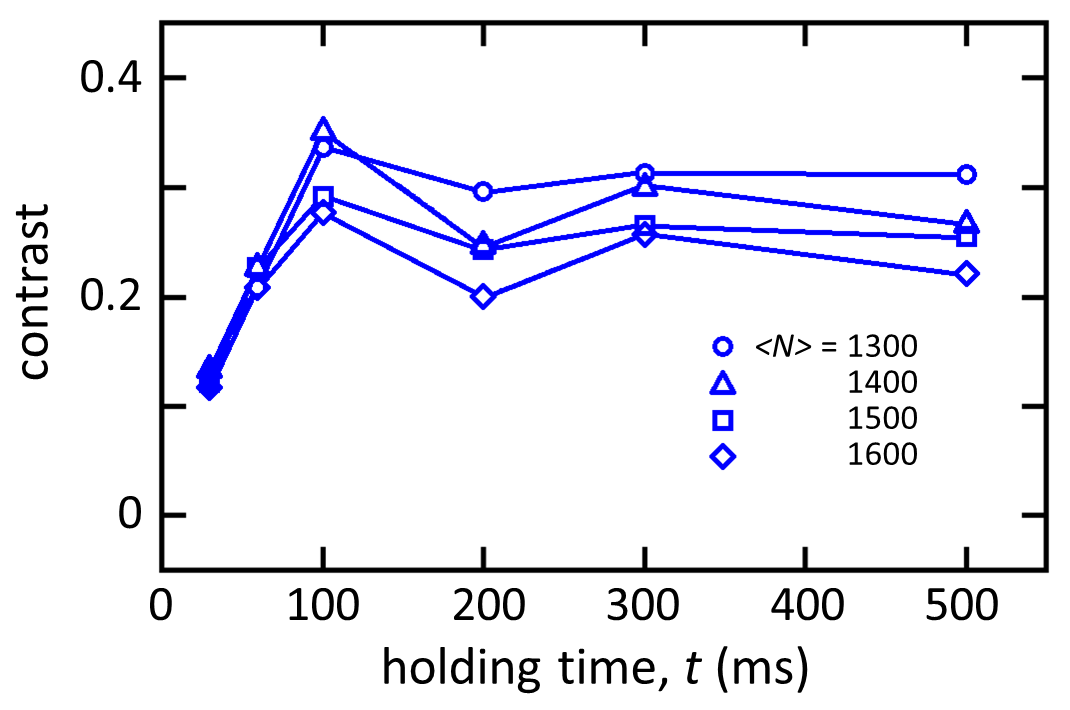}
   \caption{(Color online) Contrast \vs\ time for varying atom numbers~$N$. Here we use an alternative analysis to that of Fig.~\ref{fig:contrast}(a) by analyzing data samples that are chosen to conform to a certain mean~$N$. Each sample consists of exactly~30 experimental cycles that are closest to the means~$\langle N\rangle$ shown. While the dependence of contrast on~$N$ [as presented in Fig.~\ref{fig:contrast}(b)] is evident, it is also clear that the qualitative behavior of the data remains constant. The chosen mean values could not be taken outside the presented range since there would not be enough images for some of the holding times.}
   \label{fig:meanN}
\end{figure}

In order to further check the robustness of the results shown in Fig.~\ref{fig:contrast}(a), we re-calculate the contrast for groups of exactly~30 images from each holding time~$t$ such that the average number of atoms~$N$ is equal for all groups. The images chosen for each group were those with~$N$ closest to the mean value, generating the alternative analysis shown in Fig.~\ref{fig:meanN}. Here we observe the same qualitative behavior as in Fig.~\ref{fig:contrast}(a), even though the contrast drops with increasing~$\langle N\rangle$, thus demonstrating that the analysis of contrast \vs\ holding time is stable and credible.

\vfill\eject

In addition to the atom number~$N$, we attempted to analyze the temperature of any residual thermal background, which also appears to depend on~$t$ [Fig.~\ref{fig:1000shots}(a)]. This analysis proved fruitless however, because the background could not reliably be extracted from the individual images due to their low optical density. Moreover, it may well be that the observed broad background is due to excited modes of the~BEC caused by imperfect loading and not thermal at all. These uncertainties do not alter the qualitative observation that the contrast is robust for~$\tau_{\rm coh}\ge\rm500\,ms$, as demonstrated by both the statistical and systematic errors shown in Fig.~\ref{fig:contrast}(a) and Fig.~\ref{fig:meanN}. This qualitative observation is conserved even when the entire data set of Fig.~\ref{fig:1000shots}(a) is used (\ie\ no restriction on~$N$).

\vfill\clearpage

\bibliography{spatial_coherence_PRA}

\end{document}